# Semi-Empirical Boundary Conditions For Strong Evaporation Of A Polyatomic Gas

Petr A. Skovorodko

*Institute of Thermophysics, Prospect Lavrentyev 1, 630090, Novosibirsk, Russia*
*E-mail: almoroz@itp.nsc.ru*

**Abstract.** Semi-empirical procedure of obtaining the boundary conditions for strong evaporation of a polyatomic gas is proposed. The procedure is based on specifying the dependencies of specific momentum and enthalpy of the flow from the gas flow rate. The latter are derived from the analysis of the available data concerning the evaporation of polyatomic and monatomic gases [1,2]. The procedure predicts the exactly sonic terminal velocity of the flow and leads to some corrections of boundary conditions [1] in transonic region. Thus, for j=2 and sonic flow the corrected values of pressure, temperature and fraction of back-scattered molecules are about 1.007, 0.97 and 0.92 of those reported in Ref. 1.

## INTRODUCTION

The problem of polyatomic gas evaporation attracts much less attention in comparison with the case of monatomic gas. This circumstance seemed to be caused by more complex character of the flow of evaporated molecules with internal degrees of freedom.

For some problems of hydrodynamics and thermophysics, which are connected with evaporation of molecular vapour from solid or liquid state into flooded space, it is necessary to know the relation between the parameters on the evaporated surface and that on the external boundary of the Knudsen layer without detailed information about the structure of the layer. A good approach for these boundary conditions is proposed in Ref. 1, where the problem is treated based on the model representation of the distribution function. The analysis of these conditions shows, however, that they are not completely adequate for transonic terminal velocities – the maximum value of the gas flow rate g is predicted at the Mach numbers less than unity.

To correct the boundary conditions of Ref. 1 for transonic region of terminal velocities a simple semi-empirical procedure is proposed. The procedure is based on the dependencies of specific momentum i and enthalpy h of the flow from the gas flow rate g. The analysis of the data of Ref. 1 reveals that for any values of g the variable i is highly conservative to the number of internal degrees of freedom j and is practically the same as in available solution of the Boltzmann equation for monatomic gas evaporation [2]. This circumstance allows one to suppose that the remaining variable h may be obtained by proper estimation of contribution of internal degrees of freedom and adding it to corresponding value for monatomic gas. The procedure predicts the exactly sonic terminal velocity of the flow and leads to some corrections of boundary conditions [1] in transonic region.

## PROCEDURE AND RESULTS

It is well known that for the problem of gas evaporation from the plane surface only one parameter may be specified at infinity. Among these parameters may be pressure p, temperature T, density ρ, velocity u, or some their combination like Mach number M, gas flow rate g, etc. The two other parameters, which are needed to define the parameters of the gas flow, are to be found as a result of the solution of Boltzmann equation for considered problem. For each of the parameters specified at infinity the steady solution exists in definite region of its variation.

The problem of obtaining the boundary conditions is equivalent, therefore, to the problem of finding the dependencies i(g, j) and h(g, j), which are bound to the parameters of the gas flow by the relations:

$$u + T / u = i\,(g, j) \tag{1}$$

$$(5+j)\,T / 2 + u^2 / 2 = h\,(g, j). \tag{2}$$

The variables u and T in (1), (2) are in dimension form in the system of units where the pressure $p_s$ and temperature $T_s$ on the evaporated surface, as well as the gas constant R are equal to unity. The variable g is normalized by the value $(2\,\pi)^{-1/2}$, representing the maximum mass flux from the evaporated surface without back-scattered molecules.

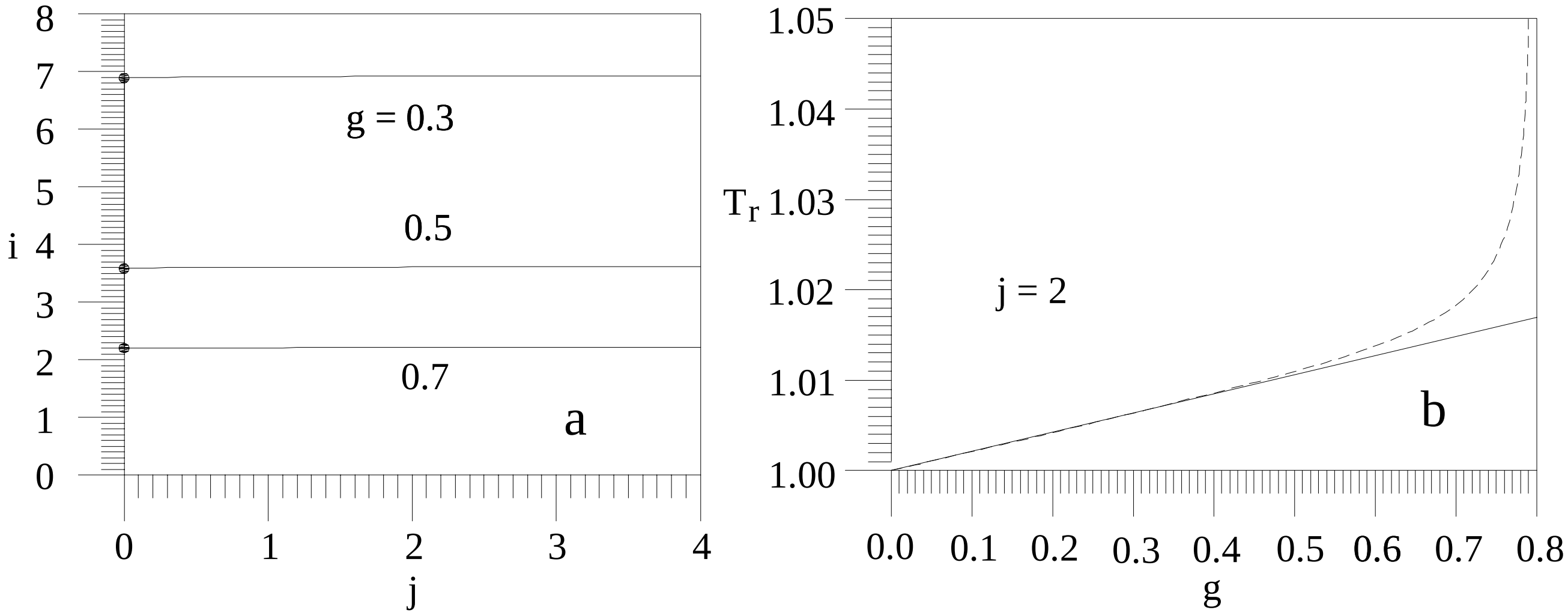

**FIGURE 1.** The dependencies of i (g, j) for g = 0.3, 0.5 and 0.7 (a) and $T_r$ (g) for j = 2 (b).

Fig. 1a represents the dependencies of i (g, j) for three values of g (0.3, 0.5 and 0.7) and $0 \le j \le 4$, predicted by the model, proposed in Ref. 1 (solid lines). The results of the solution of the Boltzmann equation for monatomic gas evaporation (j=0) [2] are also shown for comparison (solid circles). These data reveal high conservativity of specific momentum to the number of internal degrees of freedom. This circumstance is not evident *a priori* and allows one to suppose that the remaining variable h may be obtained by proper estimation of the contribution of internal degrees of freedom $T_r$ and adding it to corresponding value for monatomic gas:

$$h\,(g, j) = h\,(g, 0) + j\,T_r / 2. \tag{3}$$

Fig. 1b represents the dependence of $T_r$ (g) for j = 2 derived from the data of Ref. 1 for j = 0 and j = 2 using the relation (3) (dashed line). As can be seen from this data for the range $g \le 0.4$ the linear dependence $T_r$ (g) (solid line) takes place

$$T_r = 1 + \alpha\,(j)\,g. \tag{4}$$

Similar dependencies of $T_r$ (g) are observed for other values of j. The values of $\alpha$ (j) obtained by the least squares method for the range $g \le 0.4$ and j = 2, 3 and 4 were found to be equal to 0.02122, 0.01827 and 0.01603, respectively.

Assuming, that a sharp increase of $T_r$ (g) at large g (see Fig. 1b) is caused by approximate nature of the model [1] with prescribed form of distribution function, in the proposed procedure the relation (4) is spread on the whole range of g.

The proposed procedure consists, therefore, of combining the results of Refs. 1, 2 in the following manner. From the results of Ref. 1 the variables $\alpha$ (j) are derived, that allows one to define h (g, j) through h (g, 0) (see Eq. 3). The dependencies of h (g, 0) and i (g, 0) = i (g , j) are derived from the results of Ref. 2, which are tabulated, that is convenient for such purpose. After that the relations (1), (2) may be resolved relatively u (g) and T (g) that allows one to determine all other parameters of the flow (p, $\rho$, M, etc.). The procedure predicts the exactly sonic terminal velocity of the flow and leads to some corrections of boundary conditions [1] in the transonic region.

Fig. 2 represents the dependencies of g (M) (Fig. 2a) and T (M) (Fig. 2b) for j = 0, 2 and 3 predicted by the model of Ref. 1 (dashed lines) together with the present results for j = 2 and 3 (solid lines). The dependencies of Ref. 2 for monatomic gas (j = 0) are also shown for comparison (solid lines).

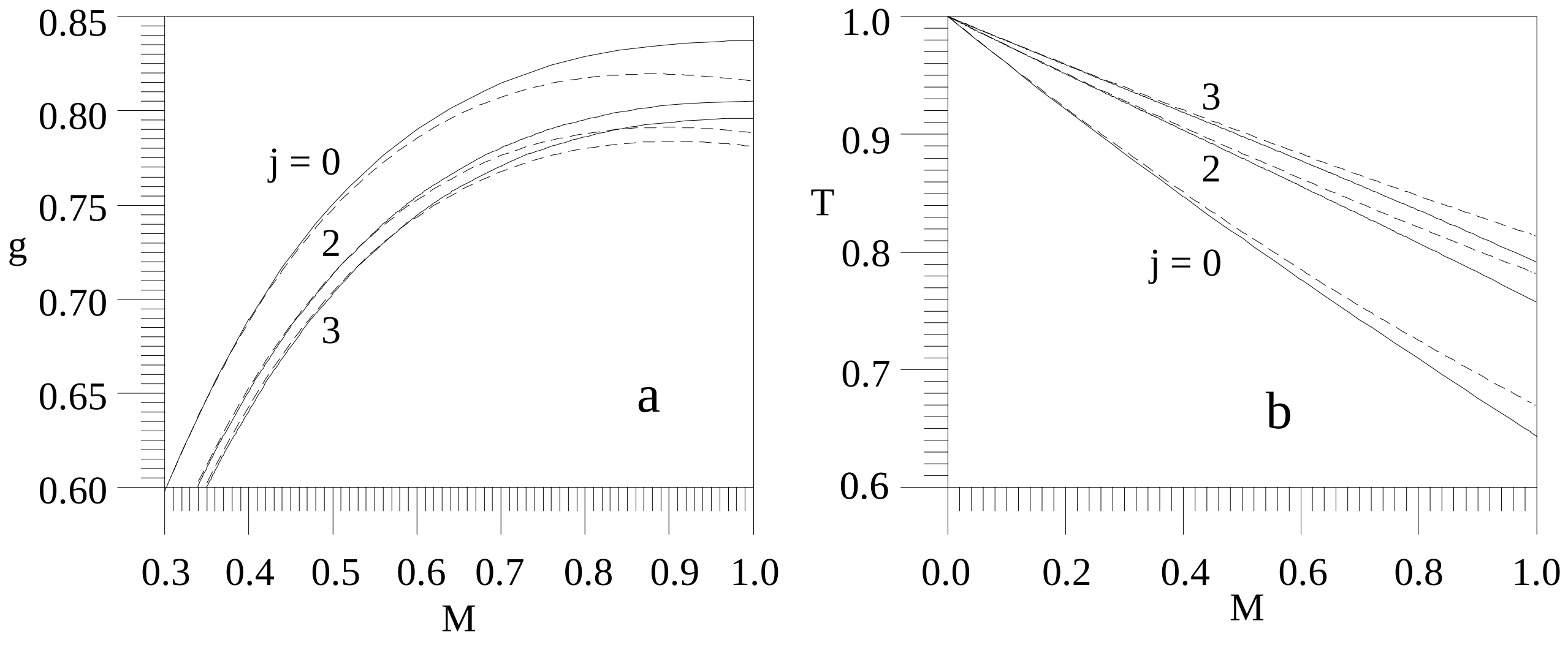

**FIGURE 2.** The dependencies of g (M) (a) and T (M) (b) for j = 0, 2 and 3.

As can be seen from the data for g (M) the proposed correction removes unphysical maximum of g at $M < 1$. The modified curves g (M) for j = 2 and 3 are higher than those from Ref. 1 for $M \geq 0.6$. These differences are qualitatively the same as the difference between the solid and dashed curves for monatomic gas.

Similar situation takes place for the dependencies T (M) (see Fig. 2b). The corrected values of T are lower than those from Ref. 1 for $M \geq 0.6$.

Table 1 contains some of the obtained results for p, T and the fraction of back-scattered molecules $\beta$ (in our notations $\beta = 1 - g$) for sonic flow (M = 1) and j = 1 ÷ 4 in comparison with the corresponding data from Ref. 1.

**TABLE 1. Boundary conditions for M = 1 and j = 1 ÷ 4.**

| j | Data from Ref. 1 | | | Present data | | |
|---|---|---|---|---|---|---|
| | p | T | β | p | T | β |
| 1 | 0.2233 | 0.737 | 0.201 | 0.2247 | 0.711 | 0.182 |
| 2 | 0.2350 | 0.781 | 0.212 | 0.2366 | 0.758 | 0.195 |
| 3 | 0.2434 | 0.814 | 0.219 | 0.2450 | 0.791 | 0.204 |
| 4 | 0.2498 | 0.837 | 0.224 | 0.2513 | 0.817 | 0.210 |

As can be seen from these data for j=2 the corrected values of pressure, temperature and the fraction of back-scattered molecules are about 1.007, 0.97 and 0.92, respectively, of those reported in Ref. 1.

## CONCLUSION

The boundary conditions for evaporation of polyatomic gas, proposed in Ref. 1 are good enough for wide range of terminal Mach number except the transonic region of velocities, where some corrections are needed. Such corrections are provided by the proposed procedure. The correction leads to lower values of temperature and fraction of back-scattered molecules, but practically to the same values of pressure. Additional investigations are needed to check the obtained results.